\documentclass[a4paper, 12pt]{article}
\usepackage{a4}
\usepackage{amsfonts}
\usepackage{amssymb}
\usepackage{cite}
\usepackage{epsfig}
\usepackage[all]{xy}
\usepackage{amsmath}
\usepackage{cite}

\author{}

\newcommand{\be}{\begin{equation}}
\newcommand{\ee}{\end{equation}}
\newcommand{\ba}{\begin{array}}
\newcommand{\ea}{\end{array}}
\newcommand{\bea}{\begin{eqnarray}}
\newcommand{\eea}{\end{eqnarray}}

\newcommand{\ov}{\overline}
\def\IR{\relax{\rm I\kern-.18em R}}

\def\IP{\relax{\rm I\kern-.18em P}}
\def\inbar{\vrule height1.5ex width.4pt depth0pt}
\def\IC{\relax\,\hbox{$\inbar\kern-.3em{\rm C}$}}

\def\K3{{\bf K3}}

\def\ov{\overline}

\def\n2d{\cN_{V^*}^{\otimes 2}}

\def\IC{\mathbb{C}}

\def\IR{\mathbb{R}}

\def\IP{\mathbb{P}}

\def\cN{{\mathcal N}}

\def\cP{{\mathcal P}}

\def\to{\rightarrow}

\begin{document}

\title{
\begin{flushright} \vspace{-4cm}
\small CERN-PH-TH/2008/086 \\
 \small LMU-ASC 30/08\\
 \small LPTENS-08/28\\
 \small MPP-2008-49 \\
\small ROM2F/2008/12 \\
 \small UPR-1195-T\\
\end{flushright}
{\bf Comments on  Orientifolds \\
without Vector Structure} }
\vspace{0.1cm}
\author{Constantin Bachas$^{\clubsuit}$,
Massimo Bianchi$^{\heartsuit , \flat}$,
Ralph~Blumenhagen$^{\spadesuit}$,\\ Dieter L\"ust$^{\spadesuit,\diamondsuit}$ and
Timo Weigand$^{\sharp}$}

\date{}

\maketitle

\begin{center}
\emph{$^{\clubsuit }$ Laboratoire de Physique Th\'eorique de l'\'Ecole Normale Sup\'erieure, \\
24 rue Lhomond, 75231 Paris,
France \footnote{Unit\'e mixte de recherche (UMR 8549)
du CNRS  et de l'ENS, associ\'ee \`a l'Universit\'e  Pierre et Marie Curie et aux
f\'ed\'erations de recherche
FR684  et FR2687.}}\\
\vspace{0.15cm}
\emph{$^{\heartsuit}$ Physics Department, Theory Unit, CERN\\
         CH1211, Geneva 23, Switzerland}\\
         \vspace{0.15cm}
 \emph{$^{\flat}$ Dipartimento di Fisica \& Sezione INFN, \\
          Universit\`a di Roma ``Tor Vergata''\\
 Via della Ricerca Scientifica, 00133 Roma, Italy}\\
\vspace{0.15cm}
\emph{$^{\spadesuit }$ Max-Planck-Institut f\"ur Physik, F\"ohringer Ring 6, \\
  80805 M\"unchen, Germany } \\
  \vspace{0.15cm}
\emph{$^{\diamondsuit }$ Arnold-Sommerfeld-Center for Theoretical Physics, \\Department for Physics,
  Ludwig-Maximilians-Universit\"at M\"unchen, \\Theresienstr. 37, 80333 M\"unchen, Germany } \\
\vspace{0.15cm}
\emph{$^{\sharp}$ Department of Physics and Astronomy, University of Pennsylvania, \\
     Philadelphia, PA 19104-6396, USA }
\vspace{0.3cm}


\end{center}

\begin{abstract}
\noindent We revisit type I compactifications with a
$Spin(32)/{\mathbb Z}_2$ gauge bundle that admits no vector
structure. We elucidate the relation of this  ${\mathbb Z}_2$
obstruction to discrete $B$-field flux and to 't Hooft flux and
clarify some subtleties in the T-duality transformation to type
IIA intersecting D-brane models. We reexamine the earliest
3-generation GUT model on magnetized D-branes and show its
consistency when a  discrete $B$-flux is switched on. We further
generalize partially known results for toroidal models to type I
compactifications without vector structure and their mirror dual
type IIA orientifolds on genuine Calabi-Yau manifolds. We
illustrate this by working out the example of the quintic in some
detail.

\vskip 3mm\
\end{abstract}

\thispagestyle{empty}
\clearpage

\tableofcontents

\section{Introduction}

The purpose of the present note is to clarify an issue arising in
the study of compactifications of  type I and heterotic string
theories.  Such  compactifications are  specified by the choice of
an internal manifold,  $X$,  and of   a gauge bundle over $X$
suitably embedded in    $Spin(32)/{\mathbb Z}_2$.
 Because the  gauge group is
$Spin(32)/{\mathbb Z}_2$,  rather than  $SO(32)$,   certain
choices which would be forbidden in the latter case are in fact
allowed. These are the gauge bundles ``without vector structure".
They play an important role  in the discussion of  various string
dualities, as pointed out some time ago in \cite{Berkooz:1996iz,
Bianchi:1997rf, Witten:1997bs} and further analyzed in
\cite{Aspinwall:1996vc, Sen:1997pm, Lerche:1997rr,
Kakushadze:1998bw, Angelantonj:1999xf, Angelantonj:1999jh,
Angelantonj:2000xf, Kakushadze:2000hm, Keurentjes:2000bs, de
Boer:2001px, Keurentjes:2001cp}. \vskip 1mm

A crucial ingredient  of the discussion on the type I side  is the
option of  turning  on a non-zero but quantized background of the
internal NS-NS 2-form $B_{ij}$, which is odd under the worldsheet
parity $\Omega$. This  was recognized early on, based on
intuition gained from rational models \cite{Bianchi:1989du,
Bianchi:1990yu, Bianchi:1990tb}, in the first systematic study of
toroidal compactifications of the type I theory
\cite{Bianchi:1991eu}.  The key observation is that since the flux
 of $B$ through any 2-cycle $\gamma$ of  $X$ is
 defined (in appropriate units)
 up to $2\pi$  shifts,  both  $\int_\gamma B = 0$ and $\int_\gamma B= \pi$ can be
compatible with the $\Omega$  projection.  These discrete
closed-string moduli are thus described by an element  of a ${\rm mod}\,2$
cohomology,  $b \in H^2(X, {\mathbb Z}_2)$. A  worldsheet argument
\cite{Sen:1997pm}  then shows  that the gauge bundle,  $E$, supported
on the D9-branes
 of the type I theory must obey the consistency condition
\be\label{b=w} b=   \tilde w_2(E) \ , \ee where  $\tilde w_2$ is a
generalized Stiefel-Whitney class which measures the obstruction
to endowing $E$ with  vector structure \cite{Berkooz:1996iz}. In
the special case of toroidal models and flat $E$, a non-zero
torsion class $b$ leads to unbroken gauge groups with reduced rank
\cite{Bianchi:1991eu}. This statement acquires  a more intuitive,
geometric meaning when translated  in the T-dual language of type
IIA orientifolds, discussed in \cite{Angelantonj:1999xf,
Angelantonj:1999jh, Angelantonj:2000xf}. Here we will clarify the
precise meaning of the consistency condition (\ref{b=w}), and
further elucidate the T-duality transformation and the reduction
of the rank. For simplicity we will perform this analysis for the
simplest case of compactifications on a single two-torus in
section \ref{GBonT2}.  This admits a straightforward
generalization to four-dimensional models on $T^2 \times T^2
\times T^2$ as will be summarized in section \ref{Torus4D}.

 \vskip 1mm

In a different development, one of us  (CB)  noted that type I
theory on magnetized tori presented  many interesting
phenomenological features, which were illustrated with   a simple
(non-supersymmetric but only marginally unstable)  grand-unified
3-generation model \cite{Bachas:1995ik}.  The  systematic analysis
of the model-building possibilities of type I magnetic fields, and
of  their T-dual intersecting D-branes, started  with the work in
\cite{Blumenhagen:2000wh, Angelantonj:2000hi,Blumenhagen:2000ea}
and has been  very
actively pursued thereafter
 (for reviews and more references see \cite{Angelantonj:2002ct,
 Uranga:2003pz, Kiritsis:2003mc, Blumenhagen:2005mu,
Blumenhagen:2006ci}).
The 3-generation model of reference \cite{Bachas:1995ik}
 (hereafter called for short  ``model C'') was actually discarded
 in \cite{Blumenhagen:2000wh},
as being T-dual to a type IIA orientifold
 with half-integer  D6-brane wrapping numbers.
 Closer inspection,  however,  reveals that the magnetic fields of model C
 describe  precisely a  $Spin(32)/{\mathbb Z}_2$ bundle without vector structure.
As we will explain in section \ref{Torus4D}, turning on the discrete $B$-field
background
 required by condition (\ref{b=w})  makes model C consistent, and restores
 the integrality of the D6-brane  wrapping numbers  in the IIA  picture.
   Furthermore, contrary to the case of flat bundles, the rank of the unbroken gauge
   group is not reduced.

\vskip 1mm Part of our motivation in writing this note was the
wish to clarify this subtle point, and to amend/rectify the
relevant statements in \cite{Bachas:1995ik, Blumenhagen:2000wh}.
Moreover, gauge bundles without vector structure have not been
much used in (semi)-realistic  model building so far.  The
simplicity of the  gauge bundle  of model C, which yields quite readily
  a 3-generation grand-unified model, is an encouragement to 
  further explore this direction.

\vskip 1mm As one step towards further  applications we generalize
the framework to type I compactifications without vector structure
on genuine Calabi-Yau manifolds in section \ref{CYmodels}. Their
mirror dual type IIA orientifolds are distinguished by allowing
for a non-vanishing real part of some of the complex structure
moduli. While this freedom is usually not much appreciated in the
literature it does amplify considerably the possibilities for model
building. We conclude by illustrating these observations for the
example of the quintic.


\section{Gauge bundles and orientifolds on $T^2$}
\label{GBonT2}

We will  first discuss gauge bundles without vector structure in
the simplest  case of a toroidal Type I compactification
\cite{Bianchi:1991eu}. Although for such backgrounds most of the
work on model-building has been already done,  our  discussion
will hopefully shed some more light on a few subtle
points\footnote{{See also \cite{Pesando:2008xt} for recent work on
the subject from a different vantage point.}}. It also serves as
preparation for the discussion of orientfiolds without vector
structure on general Calabi-Yau spaces in section 4.

 \subsection{Long strings and 't Hooft flux}

Consider compactification of  type II string theory on a
2-torus parameterized by $(x^8, x^9)\in [0,1]\times [0,1]$. The
torus is  wrapped by  a stack of $n$   D($2+k$)-branes which carry
on their world-volume a $U(n)$ gauge field $(A_8, A_9)$. In this
section the extra $k$ dimensions of the D-branes will be inert, so
we may as well set  $k=0$.
  We are interested in configurations
 \be\label{mag1}
A_8=0\ ;
 \ \ \    A_9 =  {\rm diag} [ f^{1}, \cdots ,  f^{n}]\,  x^8 +  {\rm diag} [ \alpha^{1}, \cdots ,  \alpha^{n}]\  ,
  \ee
 corresponding to a
 constant  diagonal  magnetic field $F_{89}=  {\rm diag} [ f^{1}, \cdots  f^{n}]$. This background field defines the field strength of a gauge bundle, and
the observable gauge symmetry on the D-branes is the commutant of the structure group of this bundle in $U(n)$ (modulo the issue of massive $U(1)$ factors.)
 If there are no other D-branes in the problem,   the
 first Chern class,  which counts the number of D-particles,  must be integer:
 \be\label{Chern}
m  \equiv {1\over 2\pi}   \int_{T^2}\,  {\rm tr}\, F_{89}\,   = \,
\sum_{I=1}^n {f^I\over 2\pi}    \   \in\  {\mathbb Z}\ .
 \ee
 We would like to understand the quantization conditions for  the individual $f^I$. The argument
 is well-known, but we summarize it here for completeness.

 \vskip 1mm

 If the structure group of the gauge bundle were $U(1)^n$, then standard Dirac quantization condition
 would impose that
 \be\label{quant}
  {f^{I}\over 2\pi}  = m^{I} \in {\mathbb Z}\  \ \ \ \ \forall I.
 \ee
But if we choose the structure group to fill the full $U(n)$,
 these  conditions are in
fact too restrictive. One example of an allowed gauge bundle  that
violates them  is \be\label{longD} f^{I} = {2\pi \over n} \ \ \
{\rm and}  \ \ \ \alpha^{I} =  {2\pi I \over n}\ . \ee As one can
easily verify, this is a consistent configuration because
\be\label{longDD} A_\mu (x^8+1) = {\cal U}^{-1}  \left[ A_\mu
(x^8)   + i   \partial_\mu\right]  {\cal U} \  , \ \ \ \ {\rm
with}\ \ \  {\cal U} = e^{-2\pi i \times  {\rm diag} [1 \cdots 0]
x^9 }\  \cP  \ . \ee Here $\cP$ is the cyclic-shift permutation
that sends $I\to I+1$, and the gauge transformation ${\cal U}$ is
periodic when   $x^9\to x^9+1$,   as it should be. Notice that the
transition functions ${\cal U}$ cannot be chosen in $U(1)^n$,
except when  the  fluxes $m^I$ are integer.  Other consistent
non-abelian gauge bundles, with fractional fluxes $f^I \in 2\pi
{\mathbb Z}/n$, can be constructed  in a similar way.

\vskip 1mm

\begin{figure}[ht]
\centering
\hspace{40pt}
\includegraphics[width=0.8\textwidth]{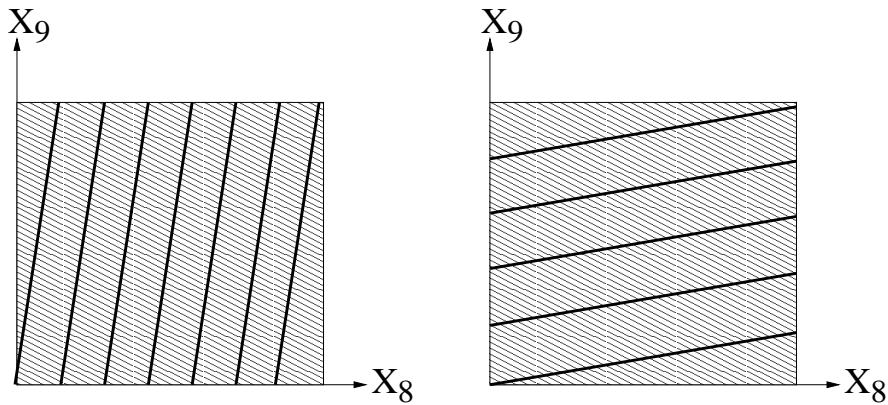}
\vspace{-10pt}
\caption{The left figure shows the T-dual of an abelian bundle
$(n,m)=(1,7)$ and the right image the T-dual of a 't Hooft bundle
$(n,m)=(5,1)$.
\label{figthooft}}
\end{figure}

These  gauge-theory statements acquire a simple geometric meaning
after a T-duality  in  the $x^9$ direction. The duality transforms
the D2-branes to D-strings  and sends  $A_9 \to  2\pi Y^9$,  where
$Y^9$ is the position of the D-strings  in the transverse dualized
dimension.
As shown in figure \ref{figthooft},
the abelian bundles (\ref{quant}) get mapped  to  configurations of $N$ independent
 D-strings with   integer winding numbers $(1, m^I)$.  Furthermore, their
  transverse positions  $\alpha^{I}/2\pi$ are unconstrained, and the unbroken
gauge symmetry has rank  $n$. The configuration T-dual to
(\ref{longD}), on the other hand,
  has $n$ pieces of D-string  with fractional winding number $1/n$ on the
dual torus. These  combine suitably so as  to form a single ``long D-string"
winding  $(n, 1)$ times  in the $(x^8, \tilde x^9)$ directions.
Such long-string configurations are familiar from the
counting  of black-hole microstates  and from the Matrix-model proposal for
M theory, see for example \cite{Ganor:1996zk,Rabadan:2001mt}.
 Notice that the above long D-string is the minimal-energy configuration
in the $(n, 1)$ topological sector.
\vskip 1mm

It is instructive for our purposes here  to separate  the
transition function  (\ref{longDD})  into a $U(1)$  phase and an
$SU(n)$ part, i.e. to write \be\label{spl}    {\cal U}  = e^{-2\pi
i x^9/n }\ e^{-i\pi/n} \, \widehat {\cal U}\ \  \ \ \ {\rm with}\
\ \ \ \ \widehat {\cal U}(x^9) \in SU(n)\ . \ee Neither of the two
factors is periodic when $x^9\to x^9+1$, but  the phases $e^{\mp
2\pi i /n}$ that they acquire cancel in the product. As a result
(\ref{longD}) cannot  be split into separately consistent $U(1)$
and $SU(n)$ bundles, but it could be separated  into consistent
$U(1)/{\mathbb Z}_n$ and $SU(n)/{\mathbb Z}_n$ bundles if there
were no particles transforming  under the center ${\mathbb Z}_n$
of $SU(n)$. In physicist's language  the latter bundle, although
flat,  carries a non-zero 't Hooft flux \cite{Hooft:1979uj},
 which is responsible for the
 breaking of the observable gauge symmetry   and the reduction of its rank from $n$ to $1$.
The  't Hooft flux is an obstruction to
  ``$n$-ality", or to ``fundamental structure"  of the  $SU(n)/{\mathbb Z}_n$ bundle,
much like   the obstruction  to vector structure which is the subject
of the present note.

\vskip 1mm

Generalizing the above example, we can define an  obstruction to
``n-ality"  for any $SU(n)/{\mathbb Z}_n$ bundle $\widehat{V}$,
whether flat or not. If this
  is part of a consistent $U(n)$ bundle, then the obstruction can be related to
  the $U(1)$ flux, encoded in the Wilson loop
\be\label{tH} {\cal W}_n(\widehat{V}, T^2) \equiv  \widehat{\cal
U}(x^9+1)\, \widehat{\cal U}^\dag(x^9) =  e^{  i \int_{T^2} {\rm
tr} F /n} = e^{2\pi i m/n}\  \in\ {\mathbb Z}_n\ . \ee
 The obstruction is thus determined by  the number of D-particles modulo the number
 of D2-branes.  Note that bundles with abelian transition functions and integer $m^I$ are
 also  obstructed whenever  $m  \not= 0 \, ({\rm mod} \, n)$.
 Note also that when $n$ is not prime  the  obstruction may concern only
a subgroup of the center ${\mathbb Z}_n$.

\vskip 1mm
In slightly more mathematical terms,
 the internal gauge fields we are considering
  correspond  to  stable $U(n)$ bundles on the torus. Stability here guarantees  that
the field strength of the associated connection is constant, i.e.
a solution of  the hermitian Yang-Mills equation.\footnote{On the
T-dual IIA side the corresponding D-strings are linear and thus
special Lagrangian.} Given such  a stable $U(n)$ bundle $V$, then
if $c_1(V)=m  \in n \, {\mathbb Z}$ we can split off a line bundle
${\cal L}$ as $V =  \widehat{V}\otimes {\cal L}$, where the
structure group of $\widehat{V}$
 is now $SU(n)$. It is known that stable $SU(n)$ bundles on a torus split
 into the direct sum of $n$ line bundles, $\widehat{V}= \bigoplus_i {\cal L}_i$ (see e.g. \cite{Friedman:1997yq}). As a result of this splitting, the rank of the visible gauge group is not reduced, and we are left with
 a $U(1)^n$ gauge theory.
 The above splitting does not, however, occur for stable $U(n)$ bundles
 with $c_1(V)$ not a multiple of $n$, or for
stable $SU(n)/{\mathbb Z}_n$ bundles.
Such bundles can, however,  be always obtained by
 deforming the direct sum of line bundles into a non-trivial extension.
  In the type IIA language, a
 piecewise-linear D-string passing
 at each step through a node of the compactification lattice can be deformed,  after
 enlarging  the structure group,  to a  linear D-string of minimal length.
This is the meaning of switching on non-trivial 't Hooft flux.

 \vskip 1mm

  In what follows we will be interested in the particular  case  of $n=32$
  D-branes,
 with first Chern class  $m  = -16$.
 From  equation (\ref{tH})  we conclude  that  the D-branes carry
 an
 $SU(32)/{\mathbb Z}_2$ bundle whose lift to a full $SU(32)$ bundle is obstructed.
 There exist two simple choices for such an obstructed bundle:
 (i) a flat bundle with  ${\mathbb Z}_2$ 't Hooft fluxes  which corresponds  to
 joining the dual D-strings in pairs,  leading   to a reduction of the rank from 32 to 16;
 or   (ii) a bundle with  half-integer magnetic fields which,  when combined
  with the $U(1)$  flux, make all the $f^I/2\pi$ integer.  The rank  in this case
  is not reduced.
  An example that illustrates
  this second  option  is
  the  abelian $U(1)^{32}$ bundle
  \begin{eqnarray}
  {F_{89}\over 2\pi}  &=&  {\rm diag}[ 0, \cdots 0,  -1, \cdots -1] \cr
&&\hskip -0.7cm
=\  -{1\over 2}\,    {\rm diag}[ 1, \cdots 1, 1, \cdots 1]  +  {1\over 2}\, {\rm diag}[ 1, \cdots 1, -1, \cdots -1] \ ,
\end{eqnarray}
   where in the first line there are 16 zeros and 16 minus ones.
Note that extracting the  diagonal $U(1)$  left us  with ``half-integer" magnetic fields
 in the remaining $SU(32)/{\mathbb Z}_2$ bundle.
Mixed configurations, with both half-integer magnetic fields and
${\mathbb Z}_2$ 't Hooft fluxes, are  also possible as we discuss
later.


 \subsection{Relation to the $B$-flux}

The above considerations  made no assumptions about the
closed-string moduli.  The NS-NS background  $B_{89}$ does,
however, affect the dynamics of the magnetized D-branes, as is
evident for instance  from the fact that the  $U(1)$ magnetic
field appears in the Dirac-Born-Infeld action only through  the
invariant combination ${\cal F} = B {\bf 1}  +  F $. This is
invariant under  the NS-NS gauge transformations
\be\label{NSgauge} B_{\mu\nu} \to B_{\mu\nu}  + \partial_\mu
\Lambda_\nu - \partial_\nu \Lambda_\mu\  \ \ \ \ \ {\rm and}\ \ \
\ \ A_\mu \to A_\mu - \Lambda_\mu\, {\bf 1}\ , \ee where the
one-form $\Lambda$ defines a $U(1)$ bundle over $T^2$ (we use here
the convention $2\pi\alpha^\prime = 1$). Large gauge
transformations change, as is well-known,  the number of
D0-branes.
 Choosing, for instance,
$ \Lambda =  2\pi\,  x^8 dx^9$  transforms $f^I \to f^I -2\pi$  and hence
 $m \to m  - n$. The first Chern class $c_1(F)$ defines therefore a quantized
 but not gauge-invariant charge.

\vskip 1mm

 A  ``physical"   D0-brane charge, which is gauge-invariant but
  not quantized,\footnote{The different notions of charge
are  even  subtler in the general case  where the NS-NS 3-form
$H=dB$  does not vanish. For a discussion see references
\cite{Bianchi:1997gt, Bachas:2000ik, Taylor:2000za,
Alekseev:2000ch, Marolf:2000cb, FigueroaO'Farrill:2000kz}.}
 can be defined as the first Chern class of the bundle ${\cal V}$ with field strength
   ${\cal  F}$,  \be\label{physQ}
 q \,   \equiv\,  {1 \over 2\pi}   \int_{T^2}\,    {\rm tr}\, {\cal  F}_{89}\,
\,  =\,   m  + {n \over 2\pi}   \int_{T^2}\,    {B}_{89}
 \  .
 \ee
 Notice that the background $B$ field induces (fractional) D0 charge on
 the D2-branes, in the same way as the
Yang-Mills $\theta$-angle  induces electric charge on magnetic
monopoles \cite{Witten:1979ey}. Suppose now that we  insist  that
the {physical}  D0-brane charge vanish. From  equations
(\ref{physQ}) and (\ref{tH}) we then conclude that 
${\cal V}$ is identified with the $SU(n)/{\mathbb Z}_n$ bundle ${\widehat V}$
(put differently the $B$ field cancels  the  diagonal-$U(1)$  part of $F$), and
that \be\label{Bfl}
{\cal W}_n(\widehat{V}, T^2) =  e^{ - i \int_{T^2} B} \ . \ee Thus
the obstruction to ``$n$-ality"  of the $SU(n)/{\mathbb Z}_n$
bundle
 is determined by the flux of   $B$,   if one insists  that  $q = 0$.
As we will argue momentarily,  this latter condition is
automatic in the type I  theory  where the D2-branes are replaced
by D9-branes and there is no R-R 8-form  to which D7-brane charge
can couple. This reasoning establishes  the formula (\ref{b=w}) of
\cite{Sen:1997pm}.

\vskip 1mm

   Before including orientifolds, let us translate these
  statements into the more intuitive T-dual language.
    Let the K\"ahler and complex structure moduli of the original 2-torus, which we
 take for simplicity orthogonal, be {}
 \be\label{moduli}
 T  =  {1\over 2\pi} \, (- {B_{89}}  +   i \ell_8 \ell_9)  \ \
  \ \ {\rm and}\ \ \  \ U =  i {\ell_8\over \ell_9}\ ,
 \ee
where  $\ell_j$  are the circumferences of the two circles.
 A T-duality along  $x^9$ exchanges  $T$ with  $U$, so that
 for non-zero $B$-field  the dual torus  is a tilted torus.  The large gauge transformations
 (\ref{NSgauge}) correspond  to  the complex structure transformations
   {} $\tilde U\to \tilde U - 1$, which   shift the  D-string winding  numbers  appropriately,
   $
  (n, m) \to (n, m - n )\, .
 $
The obstruction to $n$-ality, determined by $m$ (mod $n$),  is not affected by this shift.
The physical D0-brane charge,  $q$,  measures the (net oriented) projection
  of the D-string on the imaginary  axis of the complex plane with coordinate {}
 \be   z \equiv \, i ( \tilde x^9 -  \tilde U x^8) \ .
 \ee
  One can easily check that  the D-string with winding numbers $(n, m)$ is
 parallel to the real-$z$ axis when
 $q$ vanishes.  Roughly speaking, the rotation of the
 D-string undoes the torus tilt in this case.
 This (minimal-length)  D-string  is
 dual to  a stack of D2-branes carrying a  flat $SU(n)/{\mathbb Z}_n$ bundle.


\subsection{Including the orientifold}

We are now ready to consider the modding out by $\Omega {\cal R}$,
where $\Omega$ is the reflection of  the worldsheet coordinate $\sigma$,  and
${\cal R}$ is a ${\mathbb Z}_2$ transformation  of the
(generalized)
 target spacetime.
In  the type IIA  theory  ${\cal R}$ flips the orientation of the
2-torus, so $\tilde T^2/{\cal R}$ is one of the three open and/or
unoriented genus-1 surfaces:  the annulus, the Klein bottle or the
M\"obius strip.\footnote{Because of the action of $\Omega$ these
surfaces should not be literally thought of as the
compactification space.}
  The first two have a purely-imaginary
complex structure, whereas the third has, in our conventions,  {} ${\rm Re}\, \tilde U = - 1/2$. From
eq. (\ref{moduli}) we see that its T-dual configuration has
$B_{89} = \pi$,  so this is
the case of interest to us here. The action of ${\cal R}$ in this
case is
 \be\label{orie}
 {\cal R} z =  \bar z  \ \ \ \Longleftrightarrow  \ \ \
{\cal R} (x^8, \tilde x^9)  = (x^8, -\tilde x^9 -   x^8  ) \ .
\ee
The fixed-point surface, $x^8 = -2\tilde x^9$,
 is an orientifold 8-plane  along the connected boundary of the M\"obius strip.
It has winding numbers  $(2,-1)$  on  the doubling torus, as illustrated
in figure 2. To cancel its R-R 9-form charge we need, therefore,
to introduce D8-branes with total winding numbers $(32,-16)$.
Allowed configurations must be invariant under the action of $\Omega {\cal R}$
which is modded out.

\vskip 1mm

\begin{figure}[ht]
\centering
\hspace{40pt}
\includegraphics[width=0.8\textwidth]{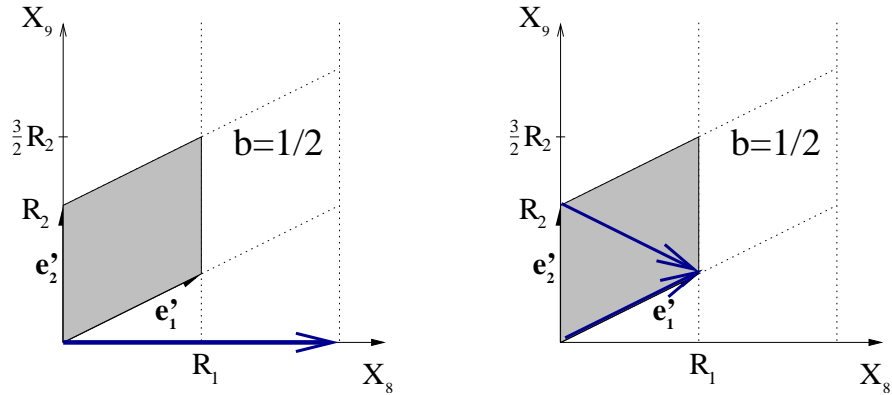}
\vspace{-10pt}
\caption{Two configurations of D8-branes canceling the
tadpole with the branes indicated by the blue arrows.
\label{figrrcancel}}
\end{figure}

Figure \ref{figrrcancel} shows two simple configurations that do the job.
The first configuration  has
 16 D8-branes along the boundary of the M\"obius strip, i.e.  with winding
 numbers $(2, -1)$ for each D8-brane.
  Because they sit on top of the orientifold, these
 D8-branes and their ${\cal R}$-images coincide. This is the supersymmetric
 vacuum,
 discussed in refs. \cite{Bianchi:1991eu, Bianchi:1997rf, Witten:1997bs}, which is
 dual to the heterotic  CHL models \cite{Chaudhuri:1995fk}. It corresponds to a flat $Spin(32)/{\mathbb Z}_2$ bundle with non-commuting Wilson
 lines \cite{Bianchi:1997rf, Witten:1997bs}
 \be
 W_1 W_2 = e^{-i\int_{T^2} B}\;  W_2 W_1.
 \ee

The second configuration in \ref{figrrcancel},  on the other hand,  has 16 D8-branes
  plus their mirror images under ${\cal R}$,  with
   winding numbers respectively
$(1, 0)$ and  $(1, -1)$.
This is the configuration on the first of the three tori of model C \cite{Bachas:1995ik}.
Standing  on its own configuration (ii) is actually  unstable,
because the $(1,0)$ and $(1, -1)$ mirror pairs can recombine  to
 form $(2, -1)$ branes.\footnote{But it
is interesting to observe that this is not allowed for a single mirror pair of D-branes.}
Model C  ``cures" this instability by exploiting the existence of the other  compactified
dimensions.

\vskip 1mm

The T-duals to the  configurations  of figure \ref{figrrcancel}
are precisely the
gauge bundles described at the end of  subsection 2.1.   We can make this
identification more
explicit by looking at the action of $\Omega {\cal R}$ on the matrix-valued
field $Y^9(x^8)$,  which  describes (in static gauge)
  the transverse position of the D8-branes.
Consistently with the geometric action (\ref{orie})  this reads
  \be\label{omega}
  \Omega {\cal R} (Y^9)  \ = \ -  \gamma_\Omega\,   (Y^9)^t   \gamma_\Omega^{-1}   - x^8
  {\bf 1} \ ,
  \ee
where  we choose (without loss of generality)   the Chan-Paton
basis so that
 \be
\gamma_\Omega  =   \left(\hskip -1mm \begin{array}{cc}  0 &  {\bf 1}_{16\times 16} \\
            {\bf 1}_{\rm\tiny{16\times 16}}    & 0  \end{array} \hskip -1mm\right)\ .
\ee
 The general solution to   the above condition is of the form
\be Y^9 = - {1\over 2} x^8 {\bf 1} + \widehat{Y}^9\ , \ee where
$\widehat{Y}^9$ takes values in the Lie algebra of $SO(32)$. Using
the T-duality dictionary,   $A_9 = 2\pi Y^9$,  one can now easily
check  that the gauge bundles of  subsection 2.1 are indeed T-dual
to those of figure \ref{figrrcancel}. \vskip 1mm

Note that although the $U(1)\subset U(32)$ gauge field is
projected out of the spectrum of the orientifold theory, a
discrete background for it actually survives. Its role is to
cancel the discrete $B$ flux so that  ${\rm tr} {\cal  F} = 0$.
Some of the confusion in the literature is due to a lack of
appreciation of this subtle point. The obstruction to vector
structure of the split-off  bundle is related  to this discrete
$U(1)$ flux and hence,  by the previous argument,   to the
discrete $B$ modulus \cite{Bianchi:1991eu}. Furthermore, the
physical D7-brane charge is
 automatically zero, consistently with the fact that the type I theory
has  no   R-R 8-form  to which this charge could couple.

\vskip 1mm

Now consider  a general configuration of D-branes which
is easier to describe in the type IIA  language.
 One  accounts
for both 't Hooft fluxes and magnetic fields by considering stacks
of D8-branes  with arbitrary  integer winding numbers. Let the
$a$-th stack have  $N_a$ D8-branes with (relatively-prime) winding
numbers  $(n_a, m_a)$.  For every stack we  must also include the
mirror stack  with winding numbers $(n_a, -m_a - n_a)$. Stacks
with  $n_a = -2m_a$ can,  a priori,  be  their own image. The
cancellation of R-R charge  requires that $n\equiv \sum_a
   N_a  n_a = 32$.
In the T-dual language
  the $a$-th stack carries a $U(n_aN_a)$ gauge bundle which has 't Hooft flux
  that  breaks the
 symmetry to $U(N_a)$, and a $U(1)$ magnetic field equal to
\be\label{magf} F_{89}^a \ =\   {2\pi m_a\over n_a}\, {\bf
1}_{N_a\times N_a}\otimes {\bf 1}_{n_a\times n_a}\ . \ee Our
normalization is such that fundamental-string endpoints have
charge $\pm 1$.
 As one can easily check, reflection
symmetry   fixes   automatically the first Chern
class of the complete $U(32)$  bundle,  as advertised  \be   m \equiv {1\over
2\pi}\, \sum_{a}  {\rm tr}\,  F^a_{89} \, = \, \sum_{a}
N_a m_a  =    -16 \ . \ee

 \vskip 1mm

 Since ${m/ n}  = -1/2$, separating the diagonal $U(1)$ gives an
 $SU(32)/{\mathbb Z}_2$ bundle $\widehat{V}$, whose lift to an $SU(32)$ bundle
 is obstructed.   Now in accordance  with the  type I
 symmetry,  the structure group of the bundle should actually be
 $SO(32)/{\mathbb Z}_2$.\footnote{We will discuss the requirement of being spin-liftable to $Spin(32)/{\mathbb Z}_2$ in section \ref{Torus1}.} 
 The reduction is
automatic if  none of the  D8-branes  is its own image. In this
case the full transition matrices  have  the block-diagonal form
\be\label{Osplit}
\widehat{\cal O}   =   \left(\hskip -1mm \begin{array}{cc}  {\cal U}  & 0  \\
           0   &  {\cal U}^*   \end{array} \hskip -1mm\right)\ ,
\ee and take values in $SO(32)$ defined as the  subgroup of
matrices that obey the reality condition $\widehat{\cal O}^* =
\gamma_\Omega\widehat{\cal O} \gamma_\Omega^{-1}$ and have
determinant 1. Such bundles can be thus  written as the sum of two
conjugate $U(16)/{\mathbb Z}_2$ bundles, $\widehat{V}  = \widehat W \oplus
\widehat W^{\vee}$.  The story is subtler for
 D8-branes which are their own  image,  and which are hence stuck to the orientifold plane.
 The elementary  ``stuck"  D8-brane has winding numbers
   $(2, -1)$  and  $\widehat{Y}^9 = 0$.
    From eqs.  (\ref{longDD}) and (\ref{spl}) we see   that the corresponding  $U(2)$
 transition function  for this D-brane reads:
 \be
  {\cal U} \, \equiv\,  e^{-i\pi x^9} \,  {\cal O} \ \ \ \ {\rm with}\ \ \ \
\,   {\cal O}  =  \,  e^{-i\pi x^9\sigma_3}\, \sigma_1\ ,
 \ee
where $\sigma_i$ are the usual Pauli matrices. The global phase in
the split-off bundle was here fixed by  imposing the reality
condition  $ {\cal O}^* = \sigma_1 {\cal O} \sigma_1$, where we think of $
{\cal O}$  as occupying the $2\times 2$  block in  the center of
the full $O(32)$ matrix.  Since  ${\rm det}\, {\cal O} = -1$, if
one insists that the full structure group be $SO(32)/{\mathbb
Z}_2$ then stuck D8-branes are not permitted.  An even number of
$(2, -1)$   D8-branes can, on the other hand, be  always combined
in mirror pairs. \vskip 1mm

 This  obstruction to the existence of a good $SO(32)/{\mathbb Z}_2$
 bundle is described by an element of a ${\rm mod}\,2$ cohomology, the first Stiefel-Whitney class
  $w_1\in H^1(X, {\mathbb Z}_2)$.  It is the same obstruction that prevents
 a pair of  $(1,0)$ and $(1,-1)$ D8-branes
 to  merge  into a $(2,-1)$ brane, as we previously  noted.
 The gauge group of \emph{perturbative} type I theory is, as a matter of fact,
 $O(32)/{\mathbb Z}_2$. It  is  non-perturbative
 consistency (see the following section) which requires that $w_1=0$,
 and  forbids\footnote{$SO(2k+1)$ factors in the gauge group are
 possible in lower dimension in the presence of `exotic' $\widetilde\Omega$-planes with
 (quantized) R-R fluxes \cite{Keurentjes:2001cp}.}
  the existence of  ``stuck"  D8-branes
 and of $SO(2k+1)$ gauge groups  in $T^2$
 compactifications \cite{deBoer:2001px}.
 In the heterotic theory the vanishing of $w_1$
 is a  perturbative requirement,  which follows from
  multiloop  modular invariance \cite{ABK}.


\section{Toroidal 4d orientifolds and the model C}
\label{Torus4D}

Now we  move on  to  more realistic backgrounds,
 obtained by
compactification of  type I theory  on
a six-torus with a non-flat  $SO(32)/{\mathbb Z}_2$
 bundle \cite{Bachas:1995ik}. These backgrounds are T-dual to
 (non-supersymmetric)  intersecting
 D6-brane models \cite{Blumenhagen:2000wh,Blumenhagen:2000ea}.
 We restrict attention to factorizable tori, $T^6 =  T^2\times T^2\times T^2$.
  For a discussion of the non-factorizable case see the
recent papers \cite{Blumenhagen:2004di,Forste:2007zb}.

\subsection{Consistency conditions for $T^6/\Omega{\cal R}$ orientifolds }
\label{Torus1}

Let  $(n^i_a, m^i_a)$ be the integer  wrapping numbers of the
$a$th stack of D6-branes on the $i$th torus,  and $(n^i_a, -m^i_a
- 2 b^i n^i_a)$ the wrapping  numbers of the mirror stack. We have
defined here  $b^i = 1/2$ or $0$, according to whether the $i$th
torus is a tilted torus or not. In the type-I language this
corresponds to a
  $B$-flux equal,  respectively,  to  $\pi$ or to  $0$.
Following  reference  \cite{Blumenhagen:2000ea}  it is also convenient to
introduce  the shifted or ``effective" wrapping numbers
 \be\label{dfn1}
  \hat m^i_a \equiv  m^i_a +  b^i\, {n^i_a}\ , \ \ \ \ {\rm so\  that} \ \ \ \
  {\cal R}(n^i_a, \hat m^i_a) = (n^i_a, -\hat m^i_a)\ \ \ \
\forall \ b^i\ .
\ee
The $\hat m^i_a$ can be considered as wrapping numbers
 along  the T-dualized  directions of the three rectangular tori,
 or as  magnetic fields
 from which  the diagonal $U(1)$ was stripped-off.
 Note that the $B$-fluxes  enter  through  the quantization conditions,
  \bea\label{hatquant}
\hat
m^i_a =  m^i_a + b^i\,  n^i_a,  \ \ \ \ {\rm where}\ \ \
 n_a^i, m^i_a \in {\mathbb Z}\ .
 \eea
For tori with $b^i = \frac{1}{2}$  the $\hat
m^i_a$ must be  integer  if  $n^i_a$ is even,  and half-integer if
$n^i_a$  is odd, while when $b^i=0$ the
$\hat m^i_a$ are always integer.  The definition (\ref{dfn1})  makes  it
  possible to treat both untilted  and tilted  tori  in a unified way.

\vskip 1mm Tadpole cancellation for the R-R 7-forms gives one
condition for each independent 3-cycle. On $T^6$ there are a
priori $20$ 3-cycles, but only 4 of them are even under the ${\cal
R}$ reflection,  $z^i \to \bar z^i$ for all $i$. One of them  is
the orientifold 3-cycle, and the other three share with the
orientifold one dimension.  The corresponding tadpole conditions
(counting branes and their images separately) read
\cite{Blumenhagen:2000wh,Blumenhagen:2000ea} \bea \label{Tadtoria}
& &\sum_{a=1}^{2K} N_a\, n_a^1\, n_a^2\, n_a^3\ =\ 32 \ ,\nonumber\\
\sum_{a=1}^{2K}N_a\, n_a^1\, \hat m_a^2\,  \hat m_a^3 &=&
\sum_{a=1}^{2K}N_a\, n_a^2\, \hat m_a^1\, \hat m_a^3 \ =\
\sum_{a=1}^{2K}N_a\, n_a^3\, \hat m_a^1\,  \hat m_a^2 \ =\ 0\ . \eea Thanks
to reflection symmetry,  tadpole cancellation for the remaining
odd cycles is automatic. Note that if the $n^i_a$ are positive,
i.e. if there are no anti-D6-branes, then maximal rank is achieved
when  $n^i_a = 1$ for all stacks and for all tori. Any  $n^i_a >1$
implies a corresponding  reduction of the rank. \vskip 1mm

Using the dictionary of the previous section,  it is  easy  to
translate the above  statements into the language of magnetized
D9-branes.  The first of the conditions (\ref{Tadtoria}) fixes the
total number of D9-branes, while the other three ensure
 that the second Chern class of the $SO(32)/{\mathbb Z}_2$ bundle on them vanishes:
 \be
 \int_{T^2_i \times T^2_j} {\rm tr} ( F^i \wedge F^j) \, =\,  0\ \ \ \ \ \forall i,j=1,2,3\ .
 \ee
These are precisely the conditions for cancellation of
D5-brane charge. The total gauge bundle has structure group $
\otimes_{a=1}^K \,  SO(2n_a)/{\mathbb Z}_2$,  where $n_a \equiv
n_a^1n_a^2n_a^3$ and we have put here stacks and image stacks in a
single  factor.
  The wrapping numbers $(n^i_a, \hat m^i_a)$ describe the 't Hooft flux and
magnetic fields of each  separate $SO(2n_a)/{\mathbb Z}_2$
bundle. Each of these bundles has an obstruction to vector structure
on the $i$th torus whenever
$b^i\not= 0$.
\vskip 1mm

Apart from R-R tadpole cancellation, additional  conditions come
from the by now recurrent observation that,  while the
perturbative gauge group is $O(32)/{\mathbb Z}_2$, the full
non-perturbative symmetry \cite{Polchinski:1995df} of type I
theory  is $Spin(32)/{\mathbb Z}_2$. Recall that $Spin(32)$ has
four conjugacy classes: $O,V,S$, and $C$, corresponding
respectively  to the adjoint, the vector, the positive-chirality
and the negative-chirality spinors. In terms of 16-dimensional
root/weight vectors these are described as follows:

\begin{table}[h!]
\centering
\begin{tabular}{ccc}
$O$: &  $(0, \ldots,0, \pm1, 0,\ldots, 0,\pm1,0,\ldots,0)$ &\\
$V$: &  $(0, \ldots, 0,\pm1, 0,\ldots,0)$  &\\
$S$: &  $(\pm \frac{1}{2}, \pm \frac{1}{2},\ldots, \pm \frac{1}{2})$\ : &even number of $+$ \\
$C$: &  $(\pm \frac{1}{2}, \pm \frac{1}{2},\ldots, \pm \frac{1}{2})$\ : & odd number of  $+$\ .  \\
\label{conj}
\end{tabular}
\end{table} \hfil \break \vskip -1.1cm
\noindent

To obtain   $SO(32)$  from $Spin(32)$ one  projects  out the spinor
representations $S$  and $C$,  while  keeping the adjoint and the
vector. Keeping only the adjoint gives the symmetry
$O(32)/{\mathbb Z}_2$.
By contrast, $Spin(32)/{\mathbb Z}_2$
keeps  the adjoint $O$ and  the positive chirality spinor
representation $S$,  while projecting out the vector representation and
the other spinor.

\vskip 1mm

     The spinor representation $S$ arises in type I theory via
  D-particle states
  which are non-BPS yet  stable  in $D=10$ and become BPS upon toroidal
  compactification
  \cite{Sen:1998tt}.
  These are dual to massive states of the heterotic string.
   Because states with the `wrong' chirality $C$
  do not exist, the parity transformation is not defined.
  Alternatively, the reduction
 of   $O(32)$  to $SO(32)$  can be traced to the existence of non-BPS D-instantons
   \cite{Witten:1998cd}.  One immediate consequence,
   encountered already in subsection 2.3,
   is that the first Stiefel-Whitney class,  $w_1$,  must vanish. As explained
  there, the complete $SO(32)/{\mathbb Z}_2$
   bundle  can then be written as  the sum of two conjugate
   $U(16)/{\mathbb Z}_2$ bundles,  $\widehat{V} =  \widehat W \oplus \widehat W^{\vee}$.

\vskip 1mm

 Such bundles are  spin-liftable
  if the standard Dirac-quantization condition for charges in $S$ is satisfied,
  i.e. if the  first Chern class  of $\widehat W$ is even. Explicitly,
  \be\label{sSW}
  \int_{T^2_{(3)}}  c_1(\widehat W)\
  = \  \sum_{a=1}^{K}   N_a  \,  n_a^1\, n_a^2 \,  \hat m_a^3
    \ \in\ 2 {\mathbb Z} \ ,
 \ee
and similarly for the other two tori. Note that the sum here runs
over all D-brane stacks, but not over  their mirrors. Condition
(\ref{sSW})  is known as the vanishing of the second
Stiefel-Whitney class which,  like the obstruction to vector
structure,  is an element of a   ${\rm mod}\,2$ cohomology,  
 $w_2(\widehat V) = c_1(\widehat W) \, {\rm mod} \, 2 \in
H^2(X, {\mathbb Z}_2)$, which obstructs the existence of spin
structure. It can be formulated as the requirement  of
cancellation of  K-theory charge \cite{Minasian:1997mm,
Witten:1998cd}
 which is stronger than the mere cancellation of R-R tadpoles.
Violation of   (\ref{sSW}) manifests itself also in the form of
 global SU(2) anomalies \cite{Witten:1982fp} in  the
world-volume theory of  probe D5-branes \cite{Uranga:2000xp}.


\subsection{Three generations and the model C}
\label{ModelC}

  For any  solution of the consistency conditions (\ref{Tadtoria}),
  subject to the quantization rules (\ref{hatquant}) and (\ref{sSW}), some of the
  most interesting observables  are the intersection numbers of stacks
  of D6-branes:
\bea I_{ab} = \prod_{i=1}^3  (n^i_a \hat m^i_b - n^i_b \hat
m^i_a)\ . \eea These determine the chiral spectrum in  the
effective four-dimensional theory, and in particular the number of
Standard-Model or GUT generations in (semi)realistic models of
this kind.  Note that the intersection numbers $I_{ab}$ are not
affected by the shift  (\ref{dfn1}), which is why the integer
winding numbers $m^i_a$ could be replaced by $\hat m^i_a$ in the
above expression.  Recall  also that in deriving the chiral
spectrum  $[ab]$ and $[a b^\prime]$, where $b^\prime$ is the mirror of
the stack $b$,
 should be considered separately.


\vskip 1mm

  Let us review now  (and sharpen a little)  the argument \cite{Blumenhagen:2000ea}  which shows
  that 3-generation toroidal models  can only exist  when  one or more of the tori are  tilted.
 We focus on the left-handed quarks, which correspond to open strings stretching between
 the color and weak-isospin stacks of D-branes (denoted here by the labels $c$ and $w$).
 To get 3 generations we need that $I_{cw} + I_{c w^\prime}=3$. 
 Generic models have  $I_{cw} = 3$ and $I_{c  w^\prime} = 0$, but because
  the $2$  and ${\bar 2}$ representations  are equivalent  we only require
 the above weaker condition. Now the mirror to the weak-isospin stack
  is obtained by flipping the
 sign of the $\hat m^i_w$, so that
 \be
 I_{cw} + I_{c w^\prime} =  -2 \prod_{i=1}^3 (n^i_w \hat m^i_c) -2 ( n^1_c\hat m^1_w n^2_c\hat m^2_w
 n^3_w\hat m^3_c + {\rm cyclic})\ . 
 \ee
This  can be odd only if some of the effective wrapping numbers are half-integers,
which  implies in turn  that at least one of the tori must be  tilted.  Bundles without
vector structure are thus unavoidable in all realistic toroidal-orientifold models.


 \vskip 1mm
   One of the nice features of   the
  3-family Grand-Unified model C is that it is obtained with
  a very  simple   choice  for the  $SO(32)/{\mathbb Z}_2$
  bundle  \cite{Bachas:1995ik}. The  choice is exhibited in table 1.

\begin{table}[h!b!p!]
\caption{The wrapping numbers of  model C.}
\begin{center}
  \begin{tabular*} {0.75\textwidth}{@{\extracolsep{\fill}}  | c   | | c   | c  | c | c | }
    \hline & & & &  \\
    stack   & $U(5)$ & $U(3)$ & $U(4)$ & $ \tilde U(4)$ \\   [2ex]
   \hline\hline & & & &  \\
    $ (n_1, \hat m_1)$  & $(1, {3\over 2})$  &  $(1, - {5\over 2})$ & $(1,  {1\over 2})$
    & $(1, -{1\over 2})$    \\    [2ex]   \hline   & & & &  \\
     $(n_2, \hat m_2)$ & $(1, {1\over 2})$  &  $(1, {1\over 2})$ & $(1, -{1\over 2})$
    & $(1, -{1\over 2})$   \\   [2ex]  \hline   & & & &  \\
    $(n_3, \hat m_3)$ & $(1, {1\over 2})$  &  $(1, {1\over 2})$ & $(1, {1\over 2})$
    & $(1, {1\over 2})$  \\  [2ex]
    \hline
  \end{tabular*}
\end{center}
\label{table1}
\end{table}


 From our previous discussion it should be  clear that the bundle admits no
 vector structure,  and requires $b^i =1/2$ on  all three tori.
 If instead the $b^i$ were zero, we would need  to multiply  the wrapping numbers $\hat m^i_a$
 by a factor 2, thereby  increasing to 24  the number of families
  \cite{Blumenhagen:2000wh}.  The construction of this model
  predated the discovery  of the non-perturbative structure of type I theory \cite{Polchinski:1995df},
  so the vanishing of the second Stiefel-Whitney class, eqs.  (\ref{sSW}), was not
   checked at that  time.  One can, however, verify that not only
   the tadpole conditions (\ref{Tadtoria}), but
     also eqs. (\ref{sSW}) are
  satisfied,  so  that the bundle C can be lifted to a fully-consistent
  $Spin(32)/{\mathbb Z}_2$ bundle.

\vskip 1mm
  The Standard-Model gauge group in model C is unified  in the $SU(5)$  group
  on  the first stack of D6-branes. There is also a horizontal $U(3)$ symmetry
  and a $U(4)\times \tilde U(4)$ hidden sector.
  One can easily check that $I_{5  3} = - 1$
  and  $I_{5 5^\prime} = 3$,   giving  three generations in the $10$
  and $\bar 5$ representations of $SU(5)$.  There is no chiral matter from strings
  between the hidden and observable stacks of D6-branes.
 The model is non-supersymmetric but free of tachyons, in appropriate regions
 of parameter space,  and it has the necessary scalar fields
 for  GUT, electroweak and
 horizontal-symmetry breaking  \cite{Bachas:1995ik}.  The pattern of supersymmetry
 breaking is also rather interesting: the gauge sector is maximally-supersymmetric
 at tree level, and the  breaking in the chiral-matter  sector  is tunable.
 The split-supersymmetry scenario  \cite{ArkaniHamed:2004fb,Giudice:2004tc}
 can be thus  implemented  in this model naturally
 (bearing in mind the usual problems of vacuum stability).
 A simple variant of model C has been, in fact,
 analyzed in this spirit in ref. \cite{Antoniadis:2006eb}. For other
 unified intersecting D-brane models see also
  \cite{Ellis:2002ci,Kokorelis:2002ns,Cvetic:2002pj,
  Axenides:2003hs,Leontaris:2005ax,Floratos:2006hs}.

 \vskip 1mm


\subsection{Euclidean D1-brane instantons}
\label{D5_T^2}

In subsection 2.3 we have discussed the origin of the obstruction
to vector structure for  the $Spin(32)/{\mathbb Z}_2$ bundle on the type I D9-branes.
It is easy to extend this   argument
to Euclidean trajectories of D-strings (or instantonic E1-branes)
 that wrap an  orientifolded two-torus with $B$-flux.
The logic is the same as before:
given $n$  branes of the above kind, we should look for  $U(n)$ gauge bundles,  $A^9(x^8)$,
that survive  the twisted orientifold  projection
   \be\label{omega}
  \Omega {\cal R} (A^9)  \ = \ \pm  \gamma_\Omega\,   (A^9)^t
   \,  \gamma_\Omega^{-1}   - 2\pi\,  x^8\,  {\bf 1} \ .
  \ee
The sign here  is $+$ for the D5-branes and $-$ for the
E1-instantons,  for reasons explained clearly  in references
\cite{small, Gimon:1996rq}. As in subsection 2.3, the general
solution of the above  condition is a ``half-integer" magnetic
field in the overall $U(1)\subset U(n)$ factor,  and  an
$O(n)/{\mathbb Z}_2$ or $Sp(n)/{\mathbb Z}_2$  bundle without
vector structure on the E1-brane, respectively on the D5-brane
world-volume.

\vskip 1mm

 The D5-branes in toroidal orientifolds are special limits of gauge bundles 
on the D9-branes, so we will not discuss them here further.
Let us consider instead in more detail 
E1-instantons wrapping the $i$th  $T^2$ factor,  for which $b^i= \frac{1}{2}$.
In the type-IIA language, the corresponding instantonic trajectories  must wrap  invariant
1-cycles of $T^2/{\cal R}$,  and all such cycles have even winding
number, $n=2k$,   in the $x^8$ direction.  This means,
when translated  in the type I language, that
only an even number of E1-branes,  with a non-trivial $O(2k)/{\mathbb Z}_2$ bundle
on their worldvolume, 
can wrap the obstructed 2-cycle. Note that,  in
 contrast to the D9-branes, the structure group for the E1-branes
  need not be reducible to $SO(2k)/{\mathbb Z}_2$. 
  Note also that the
 complete gauge group  for  the combined  system of D9- branes and  E1-branes  is
  $[Spin(32)\times O(2k)]/\mathbb Z_2$,   where the ``invisible"   $\mathbb Z_2$
  flips  the sign of the vector representations  of the two factor groups, thus
  leaving  the bi-fundamental representation $(32, 2k)$ unchanged. 
 
\vskip 1mm

The fact that a single E1-instanton
cannot wrap an obstructed 2-cycle has been observed previously in
\cite{Witten:1999eg}. This does not,  however,  mean that the multiply-wrapped
E1-branes make no contributions  to supersymmetry-protected quantities. 
Flatness of the Chan-Paton bundle is,  of course,  required 
for the instanton to be supersymmetric and thus  have a chance of
contributing to F-terms. Consider
for example $n=2$:  a flat $O(2)/{\mathbb Z}_2$ bundle with  't
Hooft flux is consistent with spacetime
 supersymmetry, and lifts half of the fermionic zero modes as 
  is evident in the T-dual ``long-string" picture. It should therefore contribute to the same quantities
 as the single ($n=1$)  E1-brane in compactifications  without  $B$-flux.
In principle, with only the zero modes corresponding to  Wilson lines
along $T^2$, such an instanton could contribute to the gauge
kinetic function on D9- or D5-branes. However, in the pure
toroidal case considered here, there exist extra zero modes related to the
transverse translations of the E1-instanton, so that such objects
rather contribute to higher-derivative F-terms.  

\vskip 1mm

E1-instantons have attracted recently much  attention,  because
they  can generate  phenomenologically desirable terms
in the effective  superpotential  of  type I
 models \cite{Blumenhagen:2006xt, Ibanez:2006da, Bianchi:2007fx, Cvetic:2007qj} .
Corrections to higher-derivative F-terms for $N=1$ vacua were
pioneered, in the
context of heterotic worldsheet instantons, in
\cite{Beasley:2005iu} and discussed in the language of D-brane
instantons in \cite{Blumenhagen:2007bn}.
A nice guide for elucidating the  type-I D-instanton calculus, in
a simpler though less realistic setting,   are the  $F^4$ threshold
corrections of maximally-supersymmetric,  $N=4$ vacua \cite{Bachas:1997xn,
Bachas:1997mc}.  The one-loop computation of these corrections on
the heterotic side is exact \cite{Lerche:1988zy}, so the
contributions of D-instantons are known. By comparing
threshold corrections in Type I models with non-zero $B$ flux and
in heterotic CHL models it is possible to verify that D-instantons
with even and odd $n$ correspond to different sectors of the
freely acting orbifold \cite{Bianchi:1998vq}. Precise agreement
between heterotic-worldsheet and E1-instanton corrections to 4-hyperini
Fermi couplings on $T^4/{\mathbb Z}_2$ has been recently demonstrated  in
\cite{Bianchi:2007rb}.


\section{Calabi-Yau compactifications of $Spin(32)/{\mathbb Z}_2$ bundles
with or without vector structure} \label{CYmodels}

So far we have analyzed the simplest case of toroidal compactification.
Much of the analysis  carries,  however,  over  to genuine Calabi-Yau spaces, 
with full $SU(3)$  holonomy, as we will discuss in this section.

\subsection{$Spin(32)/{\mathbb Z}_2$ gauge bundles}
\label{TypeIa}

Let us begin  with  type I compactifications on a general Calabi-Yau manifold $X$,
which for conceptual simplicity we assume to be smooth.  Through every 2-cycle
$\gamma \in H_2(X, {\mathbb Z})$ we may turn on integer or half-integer $B$ flux,
consistently with the $\Omega$ projection. These fluxes, and the
corresponding discrete K\"ahler moduli ${Re} (T_i)$,   are described by an element
${\cal B}  \in H^2(X,{\mathbb Z}/2)$, normalized so that ${\cal B}(\gamma) 
\equiv \int_\gamma  B/ 2\pi$. 
Of course, only the   mod2 cohomology  $b\equiv  [{\cal B}] \in H^2(X,{\mathbb Z}_2)$ describes 
physically-distinct vacua, so the number of inequivalent choices of discrete moduli  is
 $2^{h_{11}(X)}$.  
We are interested in $Spin(32)/{\mathbb Z}_2$ bundles on $X$. 
For  background material on Type I
compactifications with non-abelian vector bundles (but vanishing
$B$-flux)   we refer the reader to \cite{Blumenhagen:2005pm,Blumenhagen:2005zg}.


 \vskip 1mm

As in the toroidal case, the $Spin(32)/{\mathbb Z}_2$ bundle
defining the type I model can be constructed by first considering
a  $U(32)$ bundle $V$, \bea {V} = \bigoplus_a { V}_a^{\oplus N_a} \oplus \,
\bigoplus_a ({ V}^*_a) ^{\oplus N_a}. \eea Here ${V}_a$ denotes a $U(n_a)$ bundle
with $c_1({V}_a) \in H^2(X,{\mathbb Z})$,  
while the $*$ operation is defined
by dualizing the bundle and then twisting it with a line
bundle  ${\cal N}$,    \bea { V}_a^* = { V}_a^{\vee} \otimes
{\cal N} \quad\quad {\rm where}\quad\quad
c_1({\cal N}) = - 2 {\cal B}  \in H^2(X,{\mathbb
Z}). \ \eea
The twist bundle ${\cal N}$ accounts for the shift under the action of $\Omega$ 
which, in the toroidal case, was 
 encoded in the transformation  of wrapping numbers  $(n,m)\rightarrow
(n, -m -2b\, n)$. The structure group
$U(n_a)$  of each ${ V}_a$ is embedded diagonally into $U(n_a N_a)
\in  Spin(32)/{\mathbb Z}_2$ subject to the constraint $\sum n_a
N_a =16$. The resulting four-dimensional gauge group is given by
$\prod_a U(N_a)$ (modulo massive U(1) factors) along the lines of
\cite{Blumenhagen:2005pm, Blumenhagen:2005zg}. It follows
immediately from the above definitions that the total $U(1)$ flux
associated with $ V$ equals -16,   \bea \int_{\gamma} c_1({ V})  =
\frac{1}{2\pi} \int_{\gamma} {\rm tr}\, F=   -16\,   \eea for each
two-cycle  $\gamma \in H_2(X, {\mathbb Z})$ with half-integer
$B$-flux. Exactly as in  section \ref{GBonT2},    this guarantees
vanishing D7-brane charge on the D9-branes.

 \vskip 1mm

The advantage of working with the 
${ V}_a$  is that they are conventional bundles with integer
first Chern class.  This comes at the cost of introducing the
unusual orientifold-action operator $*$,  which is twisted by
the appearance of the $B$-flux.  Alternatively,   we can use the conventional
twist but work  with bundles whose first
Chern class can be half-integer. To this end we write
\bea
\label{cal V}
{ V}=  \widehat V \otimes {\cal L} \, , 
\eea
where the line
bundle ${\cal L}$ is such that  $c_1({\cal L}) =- {\cal B}  \in H^2(X,
{\mathbb Z}/2)$. After splitting off the diagonal $U(1)$ in this
way, $\widehat V$ represents a $Spin(32)/{\mathbb Z}_2$ bundle
given by the direct sum of  a $U(16)/{\mathbb Z}_2$ bundle and
its dual,  
 \bea \label{V} \widehat V= \widehat  W \oplus \widehat W^{\vee} \quad\quad\quad
 {\rm with}\quad\quad\quad
W = \bigoplus_a \widehat V_a^{\oplus N_a}. \eea
The generalization of the quantization condition 
eq. (\ref{hatquant})
reads
\bea
\label{quant_gen}
c_1(\widehat V_a) + n_a \,  b \in H^2(X, {\mathbb Z})\ .
\eea
For
non-zero $b$  and odd rank $n_a$, in particular,  the first Chern class  $c_1(\widehat V_a)$  
takes half-integer values. This violates the
Dirac quantization condition for the vector representation of
$SO(32)$,   so that   the bundle $\widehat V$ is  an  $SO(32)/{\mathbb Z}_2$ bundle
without vector structure.

 \vskip 1mm

For $\widehat V$ to be liftable to  $Spin(32)/{\mathbb Z}_2$, the bundle
$\widehat W$ must furthermore satisfy the
Dirac quantization condition  with respect to the spin
conjugacy class,
\bea
\label{SW1} c_1(\widehat W) = \sum_a N_a \int_{\gamma} c_1(\widehat V_a) \in 2 \mathbb Z
\quad\quad\quad \forall\ \gamma \in H_2(X,\mathbb Z).
\eea
This generalizes eq. (\ref{sSW}).
Finally, the tadpole cancellation condition for such Type I
compactifications  with  D9-branes only is given by \cite{Blumenhagen:2005pm,
Blumenhagen:2005zg}
\bea
   {\rm ch}_2(\widehat W) + c_2(T_X) = 0,
\eea where $T_X$ denotes the tangent bundle of $X$. 

\vskip 1mm

In general, one
can add D5-branes wrapping holomorphic curves  in $X$ provided the
total fivebrane class \bea W_5 = {\rm ch}_2(\widehat W) + c_2(T_X)
\eea is effective. 
Recall that a 2-cycle $\gamma$ with ${\cal B}(\gamma)=0$
can only be wrapped by $2k$
five-branes on $X$ (in the upstairs geometry).\footnote{The class $W_5$ in eq.
(40) is
the one after modding out by the orientifold action, i.e. it describes the set
of $n$ five-branes.} This yields  gauge group $Sp(k)$
in conventions where $Sp(1)=SU(2)$.
For ${\cal B}(\gamma)=1/2$  we must invoke non-trivial 't Hooft flux which
further breaks the gauge group on the five-branes. The minimal configuration on
a smooth manifold
now correpsonds to $2 \times 2$ fivebranes along $\gamma$ in the upstairs picture
where each of the two pairs carries a non-trival $SU(2)/{\mathbb Z}_2$ bundle.
This yields again gauge group $Sp(1)$ after modding out by the orientifold.

\vskip 1mm

For E1-instantons, by contrast, if ${\cal B}({\gamma})=0$ the Chan-Paton group is $O(k)$ and  no restrictions on $k$ arise.
However, absence of vector structure along a 2-cycle due to ${\cal B}({\gamma})=\frac{1}{2}$ is an obstruction 
for the appearance
of a single E1-instanton along $\gamma$
\cite{Witten:1999eg}. Here Dirac quantization would be violated
for the charged zero modes between the E1-instanton and the
magnetized D9-branes, which are discussed in Type I language in
\cite{Blumenhagen:2006xt, Bianchi:2007fx, Cvetic:2007qj}.

 \vskip 1mm

For E1-instantons to contribute to holomorphic quantities like the
superpotential or the gauge kinetic functions, they  must be of type
$O(1)$, i.e.  carry bundles satisfying $c_1(V_a)=0$.
As a result, the quantization condition \eqref{quant_gen} can only
be satisfied for even rank of the bundle. From this argument it seems to be
possible that in fact the structure established in section
\ref{D5_T^2} for E1's wrapping genus one  curves can be
generalized to for instance degree $k$ covers of isolated rational
curves. Namely, for $k$ even the quantization condition is
satisfied and  a contribution to the superpotential seems to be
possible.\footnote{Note that the degree $k$ cover can   be thought
of as the image of the map $z\to z^k$ which has two $k$-fold
branch cuts at $z=0,\infty$.}

\subsection{Smooth Calabi-Yau Type IIA orientifolds}
\label{Subsec_SmoothIIA}

Let us now discuss the mirror-dual side of 
type-IIA  orientifolds
on general Calabi-Yau manifolds. We will
identify,  in particular,  the  discrete freedom in the choice of complex
structure moduli which is dual  to the choice of orientifolds with and
without vector structure in type I.

 \vskip 1mm

Under mirror symmetry the pure world-sheet parity transformation $\Omega$
on a manifold $X$ is mapped to $\Omega{\cal R} (-1)^{F_L}$,
where ${\cal R}$
denotes an anti-holomorphic involution on the miror dual
Calabi-Yau manifold ${\cal W}$.
It acts on the holomorphic $(3,0)$ form $\Omega_3$ and
the K\"ahler two-form $J$ as
\bea
       {\cal R} : \Omega_3 \to e^{2i\theta} \, \overline \Omega_3, \quad
       {\cal R} : J \to  -J\; .
\eea
Without loss of generality we will set $\theta=0$ in what follows. The fixed point locus of ${\cal R}$ gives rise
to an orientifold $O6$-plane, whose tadpole is canceled
by the introduction of D6-branes wrapping special Lagrangian 3-cycles
on the Calabi-Yau manifold \cite{Blumenhagen:2002wn}.
These D6-branes are wrapped around homology 3-cycles $\pi_a$.

 \vskip 1mm

Let us first review the case of mirror dual to Type I
compactifications with zero $B$-field. The homology group $H_3({\cal
W},{\mathbb Z})$ splits into an $\Omega{\cal R}$ even and odd
part, $H_3({\cal W},{\mathbb Z})=H^+_3({\cal W},{\mathbb Z})\oplus
H^-_3({\cal W},{\mathbb Z})$ \cite{Grimm:2004ua}. The even part
contains real 3-cycles and the odd part completely imaginary ones.
Moreover, $\Omega{\cal R}$ exchanges the holomorphic and the
anti-holomorphic 3-forms, so that the volume form
\begin{equation}
\label{volume}{ {\rm vol}({\cal W})={i\over 8} \Omega_3\wedge \overline\Omega_3 }
\end{equation}
is anti-invariant, i.e.  $\Omega{\cal R}:{\rm vol}({\cal W})\rightarrow  -{\rm vol}({\cal W})$.
Therefore, the only non-vanishing intersections are between
3-cycles from  $H^+_3({\cal W},{\mathbb Z})$ and $H^-_3({\cal W},{\mathbb Z})$.

 \vskip 1mm

One can always find a symplectic unimodular basis $(A_I,B_I)$ of $H_3({\cal W},{\mathbb Z})$, $I=0,\ldots,h_{2,1}$,
where we take $A_0, B_i\in H^-_3({\cal W})$
and $B_0, A_i\in H^+_3({\cal W})$ for $i=1,\ldots, h_{2,1}$.
The intersection matrix for this choice of basis has the simple
form $A_I \cap B_J=\delta_{IJ}$ with all other intersection numbers vanishing.
Note that this defines a Poincar\'e dual basis $(\alpha_0, \beta_i)$ of $H_-^3({\cal W},\mathbb Z)$ and $(\beta_0, \alpha_i)$ of $H_+^3({\cal W},\mathbb Z)$
such
that
\bea
\label{uni-basis}
\int_{A^I} \alpha_J = \delta_{IJ}, \quad \quad
\int_{B^I} \beta_J = -\delta_{IJ}\, ,\quad (I,J=0,\dots ,
h_{2,1}).
\eea
In this basis, the holomorphic three-form $\Omega_3$ is expanded as
\bea
\Omega_3 =  \sum_I X_I \alpha_I- \sum_J F_J \beta_J
\eea
in terms of the periods
\bea
X_I = \int_{A_I}\Omega_3, \quad\quad\quad F_J =  \int_{B_J}\Omega_3.
\eea
Special geometry of the complex structure moduli space implies
that the periods along $B_I$ can be expressed as
derivatives of the prepotential  
${\cal F}(U_i)$, where one defines the quotient of two periods
\bea
U_i={X_i\over X_0}={\int_{A_i} \Omega_3\over \int_{A_0} \Omega_3}\; .
\eea
In terms of these one has
\bea
 {\partial {\cal F}\over \partial U_i}={\int_{B_i} \Omega_3\over \int_{A_0} \Omega_3}\; ,
 \quad\quad
{\cal F}_0 \equiv 2{\cal F} - \sum_i U_i {\partial {\cal F}\over \partial U_i} ={\int_{B_0} \Omega_3\over \int_{A_0} \Omega_3}\;.
\eea
The $U_i$ indeed transform under the orientifold action as
$\Omega{\cal R}: U_i\to -\overline U_i$, and a consistent choice is $Re (U_i)=0$.

 \vskip 1mm

As in \cite{Blumenhagen:2004xx} we expand the
3-cycles of the branes and the orientifold planes as
\bea
\label{def_wrap}
 \pi_a &=&\sum_{I=0}^{h_{2,1}} ( q_{a,I}\, A_I -p_{a,I}\, B_I), \quad\quad
            \pi_{{\rm O}6} = {1\over 2} \left( L_0 B_0 +
               \sum_{i=1}^{h_{2,1}}  L_i\, A_i \right).
\eea
The image brane has the expansion
$\pi'_a = -q_{a,0} A_0 -p_{a,0} B_0 +
    \sum_{i=1}^{h_{2,1}} ( q_{a,i}\, A_i + p_{a,i}\, B_i)$.
For a supersymmetric brane configuration the NS-NS tadpole
cancellation condition takes the simple form
\bea
\label{tadpole}
    -\sum_a N_a\, p_{a,0} {\cal F}_0 + \sum_{a,i} N_a\, q_{a,i}\,U_i \, 
    = \left( L_0 {\cal F}_0 + \sum_i L_i\, U_i \right),\,\, 
\eea
where the terms of zero   and   second  order in the $U_i$ cancel
due to the image branes. Equation \eqref{tadpole} encodes
 $h_{2,1}+1$ independent
conditions on the wrapping numbers of the D6-branes.

 \vskip 1mm

Mirror duality maps this type IIA orientifold with intersecting
D6-branes and O6-plane to type I= type IIB/$\Omega$
compactifications with magnetized D9-branes. The type IIB
K\"ahler moduli  {} $T_i=-b_i+iJ_i$ are defined by expanding   {} $ T=  - {\cal B} 
 + iJ$ as  {} $\sum_i ( - b_i + i J_i)\, \omega_i = T^i\, \omega_i$, where
$\omega_i$ denotes a basis of $H^2(X, {\mathbb Z})$. The
 mirror map exchanges the IIB moduli $T_i$ with the
IIA complex structure moduli $U_i$ \cite{Candelas:1990rm,Candelas:1990qd}. The above
choice $Re(U_i) =0$ is obviously  dual to $b_i = 0$ on the type I
side. Recall that in the type I case the possibility of
half-integer NS-NS flux results from the periodic identification
 $ {\cal B}(\gamma)  \simeq {\cal B}(\gamma)  + 1$, which, together with
${\cal B} \rightarrow -{\cal B}$ under $\Omega$ allows for  $ {\cal B}(\gamma)  = 0$ or 
$\frac{1}{2}$ \cite{Bianchi:1991eu}.
By mirror symmetry also the complex structure
moduli $U_i$  enjoy a shift symmetry  {} $U_i\simeq U_i-1$, so that the two
discrete values {} $Re (U_i)=0,-{1\over 2}$ are allowed.
The value {} $Re (U_i)=-{1\over 2}$ is the mirror dual of the
type I orientifold without vector structure.

 \vskip 1mm

Note that the value $Re(U_i)$ is  still measured with respect to
the old unimodular basis (\ref{uni-basis}). In general, with the tilt in the complex
structure, this basis ceases to take values in $H^3({\cal W},
{\mathbb Z})$, but rather is defined only in $H^3({\cal W},
{\mathbb Q})$.
Of course one can now define a new basis
of  $H^3({\cal W}, {\mathbb Z})$,  \footnote{For the toroidal case, this basis would be the one constructed from the fundamental cycles $e'_1$ and $e'_2$ in figure \ref{figrrcancel}.}  but this basis does not split
into even and odd parts under $\Omega {\cal R}$.

As on the torus one can  choose to
keep the nice transformation properties of the basis
(\ref{uni-basis}) and formally expand the three-cycles wrapped
by the D-branes as in (\ref{def_wrap}). The so-defined
wrapping numbers are subject to certain
constraints which ensure that the object $\pi$ is a bona fide
cycle.

 \vskip 1mm

To find the correct description we use the fact that mirror
symmetry exchanges the central charges of a B-type brane carrying
a holomorphic bundle $V_a$ on a Calabi-Yau $X$ and the dual A-type
brane wrapping the sLag $\pi_a$ on the mirror manifold
${\cal W}$. Recall that the central charges are defined as \cite{Aspinwall:2004jr}
\bea \label{centralZ}
Z_B = \int_X e^{{\cal B} - i J} \, {\rm ch}(V_a) \, \sqrt {Td(T_X)},
\quad\quad Z_A = \int_{\pi_a} \Omega_3. \eea
The expression for $Z_B$ depends on the gauge field and the $B$-field only via the gauge invariant combination
${\cal F} = F+B$. \footnote{To comply with the convention used in the discussion of toroidal models 
we have chosen ${\cal F} = F+B$ (as opposed to $F-B$) to be the gauge invariant combination. Consequently we have defined $Z_B$ in terms of $e^{{\cal B} - i J}$,
 rather than  $e^{-({\cal B} + i J)}$ as in \cite{Aspinwall:2004jr}.}

One now expands $Z_B$ and $Z_A$
along $H^2(X, {\mathbb Z})$ and $H^3({\cal W})$ and
uses the mirror map between $T_i$ and $U_i$ to express the
'wrapping numbers' $p_I$ and $q_I$ of the A-brane in terms of the
topological data of the mirror dual bundle and the Todd class of
$X$. To work this out  explicitly requires the form of the prepotential
${\cal F}(U_i)$.
As already anticipated one has 
here two choices: either  work  with the bundle $V_a$ as in eq. (\ref{centralZ}) with $c_1(V_a) \in H^2(X,{\mathbb Z})$. The corresponding wrapping numbers for the mirror dual A-brane will then be the ones with respect to the tilted basis 
taking values in $H_3({\cal W}, {\mathbb Z})$. Alternatively one absorbs the $B$-field into the gauge bundle by writing
\bea
Z_B= \int_X e^{- i J} \, {\rm ch}(\widehat V_a) \, \sqrt {Td(T_X)}.
\eea
This will give us the
effective 
 fractional wrapping numbers along the unimodular basis valued in  $H^3({\cal W},{\mathbb Q})$.

Let us treat the  two cases in turn. First, one can expand $Z_B$ as
\cite{Douglas:2006jp} \bea Z_B = Q^6 - T\, Q^4 + \frac{1}{2} T^2
\, Q^2 - \frac{1}{6} \, T^3 \, Q^0 \eea with \bea
&& Q^0 = {\rm rk} ({\cal E}), \quad \quad Q^2=   c_1({\cal E}) , \quad\quad Q^4 = {\rm ch}_2({\cal E})+ \frac{rk ({\cal E})}{24}\, c_2(T_X), \nonumber\\
&& Q^6 =  {\rm ch}_3({\cal E})  + \frac{1}{24}\, c_1({\cal E})\,
c_2(T_X). \eea where ${\cal E}$ collectively denotes the bundles
$V_a$ or $\widehat V_a$, depending on whether we absorb the
$B$-flux in the field $T=-b+iJ$ or the gauge bundle.

The analogous expansion for $Z_A$ reads \bea Z_A = \int_{\pi}
\Omega_3 = X_0 ( q_0 + \sum_i q_i U_i - \sum_i p_i \frac{\partial
{\cal F}}{\partial U_i} - p_0 {\cal F}_0). \eea Now one uses the
mirror map to identify $T_i$ with $U_i$. In the large volume limit
the prepotential ${\cal F}(T)$ takes the form \bea {\cal F}(T) = -
\frac{1}{6} T^3 + \frac{1}{2} A \, T^2  - \frac{1}{24}\, c_2(T_X) \,
T. \eea This classical expression receives worldsheet instanton
corrections away from the large volume limit. For a discussion of
the terms linear and quadratic in $T$, which do not enter the
tri-linear couplings, we refer e.g. to \cite{Hosono:1994av}. Using
this result and comparing the expansions of $Z_B$ and $Z_A$ leads
to \cite{Douglas:2006jp} \bea \label{wrapNum_gen}
&&  (p_a)_0 = {\rm rk}({\cal E}),  \quad\quad \sum_i (p_a)_i \, \omega_i =  c_1({\cal E}), \\
&&  q_0 =  {\rm ch}_3({\cal E}),   \quad\quad \sum_i (q_a)_i \, \widetilde
\omega_i = - \left({\rm ch}_2({\cal E}) + \frac{{\rm rk}({\cal E})}{12}
c_2(T_X) \right) + c_1({\cal E}) A \nonumber. \eea
Here $\widetilde \omega_i$ are the
elements of $H^4(X, {\mathbb Z})$ dual to $\omega_i$.

Again, the wrapping numbers with respect to the tilted geometry
with  $Re(U_i)=\frac{1}{2}$ correspond to ${\cal E}=V_a$. Note
that even in this case, with the overall normalization chosen, the
quantities $q_I$ need not be integer-valued even though they are
integer on
$T^2 \times T^2 \times T^2$. By contrast, if we stick to
the unimodular basis  (\ref{uni-basis}), we insert ${\cal
E}=\widehat V_a$, and obviously even the corresponding $p_i$ can
be half-integer. This generalizes the effective wrapping numbers
constructed from the elementary winding numbers $(n_i,\hat m_i)$
for $T^2 \times T^2 \times T^2$ as described in the appendix.

 \vskip 1mm

The structure  presented in this section is rather formal.   
For a concrete Calabi-Yau manifold and a specified anti-holomorphic
involution,   finding  the nice symplectic
basis used in this section is not an easy task. To really see that
these two discrete choices in the complex structure moduli space
are indeed possible, we will now 
 discuss one non-trivial example in some more
detail.

\subsection{Example: The Quintic}

While in  appendix A we will  provide some details on the straightforward
example of a toroidal orientifold, here we would like
to discuss for the simplest genuine Calabi-Yau, i.e. the Quintic, how
the framework summarized in the last section actually applies.

 \vskip 1mm

We consider the type I string compactified on the quintic,
i.e. ${X}=\IP_4[5]$, which has Hodge numbers
$(h_{21},h_{11})=(101,1)$ and whose complexified
K\"ahler modulus we denote as {}  $T=-{\cal B}+iJ$.
On the dual side we get a type IIA orientifold on the  mirror
manifold ${\cal W}=\IP_4[5]/\mathbb Z_5^3$.
The sole complex structure modulus $\psi$ is visible in the
general form of the hypersurface constraint surviving
the $\mathbb Z_5^3$ orbifold
\bea
     Z_1^5+Z_2^5+Z_3^5+Z_4^5+Z_5^5  - (5\psi)\, Z_1\, Z_2\, Z_3\, Z_5\, Z_5 =0\; .
\eea By a coordinate transformation like  $z_1\to \alpha z_1$ with
$\alpha=\exp(2\pi i/5)$ one sees that $\psi$ and $\alpha\psi$
define equivalent manifolds, so that  only the cone $0\le \arg
(\psi) < 2\pi /5$ respectively  $z=(5\psi)^{-5}$ are good
coordinates on the complex structure moduli space. The fundamental
region for $\psi$ gets further reduced by dividing by more general
coordinate transformations \cite{Candelas:1990rm}. We assume that
the type IIA anti-holomorphic involution acts just by
complex conjugation  ${\cal R}: Z_i\to  \overline Z_i$, so
that  the two half-lines $\arg (\psi)=0, \pi/5$ are the two
 real one-dimensional components of the complex structure moduli space of
the orientifold model.

 \vskip 1mm

In order to see how this is related to the discrete choices of the
$B$-field in the mirror dual type I description, we need to know
the mirror map. Luckily, for the quintic this map is explicitly
known and we just need to copy and interpret the results
\cite{Candelas:1990rm,Candelas:1990qd}.

\begin{figure}[ht]
\centering
\hspace{40pt}
\includegraphics[width=0.8\textwidth]{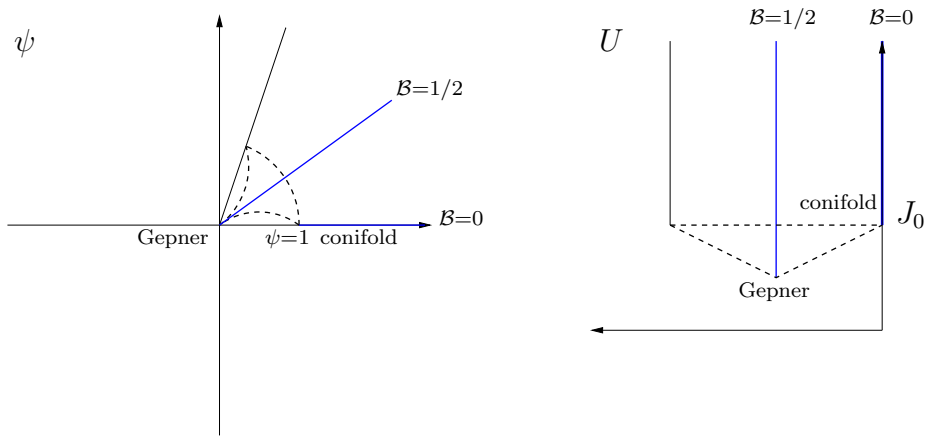}

\begin{picture}(100,1)
\put(-90,160){$\psi$}
\put(4,89){$_{\psi=1 \ {\rm conifold}}$}
\put(-44,89){$_{\rm Gepner}$}
\put(69,96){$_{{\cal B}=0}$}
\put(53,145){$_{{\cal B}=1/2}$}

\put(129,160){$U$}
\put(240,95){$J_0$}
\put(204,102){$_{\rm conifold}$}
\put(181,69){$_{\rm Gepner}$}

\put(230,172){$_{{\cal B}=0}$}
\put(185,172){$_{{\cal B}=1/2}$}

\end{picture}

\vspace{-10pt}
\caption{Complex structure moduli space for the mirror quintic ${\cal W}$
5 in the $\psi$- and the $U$-plane.
The blue lines indicate the two discrete branches
after the orientifold projection, related to ${\cal B}=0,1/2$
in the mirror dual Type I model.
\label{figquint}}
\end{figure}

In the region $|\psi|>1$, $T$ is mapped to a quotient
of periods
\bea
    U={\Phi_1\over \Phi_0}
\eea with the periods solving the Picard-Fuchs equation given by
\bea
   \Phi_0=\sum_{n=0}^\infty  { (5n)!\over n!^5} {1\over (5\psi)^{5n}}, \quad
  \quad \Phi_k=-{5\over (2\pi i)^k} \left[\log (5\psi)\right]^k\, \Phi_0 +
   \tilde\Phi_k ( \psi )\quad\quad {\rm for}\ k=1,2,3\; ,\nonumber
\eea
where, like $\Phi_0$, $\tilde\Phi_k ( \psi )$ is  an infinite series
in the variable $\psi^{-5}$.
The complex structure modulus $U$ can eventually be expressed
in terms of $\psi$ as
\bea
    U=-{5\over 2\pi i}\left[ \log (5\psi) -
          {1\over \Phi_0}\sum_{m=0}^\infty {(5m)!\over (m!)^5\, (5\psi)^{5m}}
       \left( \Psi(1+5m)-\Psi(1+m) \right)\right],
\eea where $\Psi(x)$ denotes the digamma function. Now it is clear
that $\psi\simeq \psi\, e^{2\pi i N/5}$ is mapped to the
periodicity {}  $U\simeq U-N$ and that ${\cal R}: U\to
-\overline U$. In addition, the half line $\arg (\psi)=0$ is
mapped to $T=U=i J$ with $J\ge J_0\simeq 1.21$. The other
half-line $\arg (\psi)=\pi/5$ is mapped to {} $T=U=-1/2+i J$. Note
that $\psi=1$ resp. $U=iJ_0$ is a singular point in the complex
structure moduli space, where the Calabi-Yau manifold develops a
conifold singularity. To describe the other side of the singular
point, i.e. in the region $|\psi|<1$, one is analytically
continuing  the periods to  this region. Note that in the mirror
dual type I model this region corresponds to the Landau-Ginzburg
phase of the linear sigma model. In the region around the Gepner
point $\psi=0$ the mirror map has the following expansion {}
\bea
    U=-{1\over 2} +{i\over 2}\left[ \cot\left({\pi\over 5}\right)+
        {\Gamma^4\left( {4\over 5}\right)\Gamma\left( {2\over 5}\right)\over
            \Gamma\left( {1\over 5}\right)\Gamma^4\left( {3\over 5}\right)}
           \left(\cot\left({\pi\over 5}\right)-\cot\left({2\pi\over 5}\right)
               \right)e^{\pi i\over 5}\psi + O(\psi^2) \right]\!.
\eea Suppressing a discussion of branch cuts and of the
fundamental region of $\psi$, which can be found in the literature
\cite{Candelas:1990rm}, we realize that the Gepner  point $\psi=0$
corresponds to {} $T=U=-{1\over 2} +i \cot {\pi\over 5} $. Therefore,
the Gepner point lies on the ${\cal B}=1/2$ branch, i.e. in the Type I
model it is on the same branch in K\"ahler moduli space as the
orientifolds without vector structure. The structure of the moduli
space is shown in figure \ref{figquint} (essentially taken from
 \cite{Aspinwall:1994ay}, see also \cite{Brunner:2004zd}).

For the model discussed here, i.e. the Type I string on the quintic resp.
the Type IIA orientifold on the mirror quintic, the Gepner model
orientifold was first discussed in \cite{Blumenhagen:1998tj} and featured
a maximally rank tadpole canceling solution with
gauge group $SO(20)\times SO(12)$.

\section{Outlook}

In this paper we have reconsidered Type I compactifications
without vector structure. We have offered several equivalent
descriptions that  clarify  some longstanding puzzles. In
particular we have shown the consistency of a 3 generation
non-supersymmetric but tachyon-free GUT model proposed by one of
us (C.B.) \cite{Bachas:1995ik} long time ago. The
possibility of relating ``half-integer'' wrapping numbers in the
Type IIA orientifold description to a quantized NS-NS B-field
opens new possibilities for model building and suggests a
re-analysis of toroidal compactifications with oblique fluxes
\cite{Antoniadis:2004pp, Bianchi:2005yz, Antoniadis:2005nu,
Bianchi:2005sa} in the perspective of stabilizing off-diagonal
moduli. Their mirror Type IIA description would require
``co-isotropic'' D-branes, {\it i.e.} wrapped rotated D-branes
supporting non trivial magnetic fields associated to bundles
with(out) vector structure \cite{Font:2006na,Anastasopoulos:2006hn}.

We have not explicitly considered models with different kinds of
oppositely charged but mutually supersymmetric orientifold planes
\cite{Sugimoto:1999tx, Hanany:2000fq, Bergman:2001rp,
Dudas:2001wd} that lead to models without D-branes dual to Type II
models with massive R-R sector \cite{Vafa:1995gm,
Angelantonj:1996mw}. Though an interesting playground in string
dualities \cite{Bianchi:1998vq}, at first sight this kind of
models are less appealing because of the very low rank of the
gauge group and the related difficulty in accomodating chiral
fermions. Although model C is non-supersymmetric, yet it
can be made non tachyonic by displacing the mutually non
supersymmetric stacks along the directions where they are
parallel. Moreover, one can still envisage the possibility of
introducing stacks of magnetized branes mutually supersymmetric in
pairs but not sharing any common global susy as a whole, see e.g.
\cite{Kokorelis:2002ns, Axenides:2003hs, Floratos:2006hs,
Emparan:2006it}.

 The presence of non globally
supersymmetric magnetic fields mimics the presence of lower
dimensional D-branes with opposite R-R charges and may greatly
help relaxing the stringent tadpole conditions on the rank of the
Chan-Paton group\footnote{As a `caricature' consider an (alas
tachyonic) Type I model in $D=10$ with $N+16$ D9-branes and $N$
$\overline{\rm D9}$-branes with chiral fermions and gauge symmetry
`enhancement'.} and allow for further interesting lines of
investigation.

\vskip 1cm
 {\noindent  {\Large \bf Acknowledgements}}
 \vskip 0.5cm
\noindent
We thank C.~Angelantonj, M.~Axenidis,
V.~Braun, T.~Brelidze, M.~Cveti{\v c}, J.~Evslin, E.~Floratos, A.~Klemm, 
C.~Kokorelis, R.~Minasian,
R.~Richter, A.~Sagnotti and C.~Timirgaziu for useful conversations. This work
has been supported in part by the European Community Human
Potential Program under contracts MRTN-CT-2004-005104 and
MRTN-CT-2004-512194, by the INTAS grant 03-516346, by MIUR-COFIN
2003-023852, by NATO PST.CLG.978785, by DOE grant EY-76-02-3071 and by
the Excellence Cluster
``The Origin and the Structure of the Universe'' in Munich.
C.B. thanks the
Arnold-Sommerfeld-Center in Munich, 
M.B. thanks the  Ecole Normale Sup\'erieure, 
R.B. thanks the University of Bonn, D.L. thanks the University of Pennsylvania and
T.W. thanks the University of Wisconsin, Madison, for hospitality during part of this work.

\begin{appendix}
\section{Toroidal Example}

In this appendix we demonstrate the observations of section \ref{Subsec_SmoothIIA} for the simple example of compactifications on
 $T^2\times T^2\times
T^2$. Here we have 8 homology 3-cycles
\bea
\label{untiltbasis}
& &  A_0=(0,1)\otimes  (0,1)\otimes (0,1), \quad\quad\quad \,\, \,  B_0= (-1,0)\otimes (-1,0)\otimes (-1,0), \quad \nonumber\\
& &  A_1= (-1,0)\otimes (0,-1)\otimes (0,-1), \quad  B_1=(0,-1)\otimes  (-1,0)\otimes (-1,0), \quad  \\
& &  A_2= (0,-1)\otimes (-1,0)\otimes (0,-1), \quad  B_2=(-1,0)\otimes  (0,-1)\otimes (-1,0), \quad   \nonumber\\
& &  A_3= (0,-1)\otimes (0,-1)\otimes (-1,0), \quad  B_3=(-1,0)\otimes (-1,0)\otimes (0,-1). \quad \nonumber \eea

They satisfy $A_I \cap B_J = \delta_{IJ}$. We also introduce the dual basis
$(\alpha_I, \beta_J)$  with $\int \alpha_I \wedge \beta_J = \delta
_{IJ}$,

\bea \label{Basis^3}
& &\alpha_0 = dy^1 \wedge dy^2 \wedge dy^3, \quad\quad\,\, \, \, \,  \beta_0 = dx^1 \wedge dx^2 \wedge dx^3, \nonumber \\
&& \alpha_1 = - dx^1 \wedge dy^2 \wedge dy^3, \quad\quad
\beta_1 = dy^1 \wedge dx^2 \wedge dx^3, \\
&& \alpha_2 =  -dy^1 \wedge dx^2 \wedge dy^3, \quad\quad
\beta_2 = dx^1 \wedge dy^2 \wedge dx^3, \nonumber \\
&& \alpha_3 = -dy^1 \wedge dy^2 \wedge dx^3, \quad\quad
\beta_3 = dx^1 \wedge dx^2 \wedge dy^3 \nonumber.
\eea

The orientifold plane is chosen along the $x$-direction in each $T^2$ so that indeed
$A_0, B_i \in H^-_3(T^6, {\mathbb Z})$ and $B_0, A_i \in H^+_3(T^6, {\mathbb Z})$.
The holomorphic coordinates
\bea
dz^i =- U_i dx^i +  dy^i, \quad\quad d\ov z^i = - \ov U_i dx^i +  dy^i
\eea
are determined by the complex structure moduli $U^i$.
We take $U_i = (- b_i + i u_i)$ with $u_i=\frac{R_x^i}{R_y^i}$ in terms of
the radii of the elementary 1-cycles.

In the symplectic basis (\ref{Basis^3}), the holomorphic
three-form  $\Omega_3 = dz^1 \wedge dz^2 \wedge dz^3$  enjoys the
expansion
\bea \Omega_3 =  \alpha_0  +  \sum_{i=1}^3 (U_i) \alpha_i   + \frac{1}{2}   \sum_{i\neq k \neq j}  \,
(U_i U_j) \beta_k - U_1 U_2 U_3  \beta_0.
\eea
Note that indeed the ratio of periods $\frac{\int_{A_i}\Omega_3}{\int_{A_0}\Omega_3} = U_i$.
The orientifold rule $\Omega {\cal R}:  U_i \rightarrow \ov U_i$ together with the
identification $U_i \simeq U_i -1$ translate into \bea U_i= -
\ov U_i - n. \eea Indeed, the values  $U_i=i$ and  $U_i=i-
\frac{1}{2}$ of the untilted and tilted case satisfy this with
$n=0$ and  $n=1$, respectively.

One way to describe consistent 3-cycles on the torus is by
introducing  effective wrapping numbers $q_I, p_I$ as in equ. (\ref{def_wrap}) with respect to the \emph{untilted} basis \ref{untiltbasis}, which, for $b_i =1/2$, takes values only in $H^2(T^6, {\mathbb Q})$.
For factorizable branes these are given in terms of the wrapping numbers along the horizontal and vertical axes, $n_i$ and $\tilde m_i = m_i + b_i n_i$, by
\bea
\label{wrap_T}
&& p_0 = n^1 n^2 n^3, \quad p_1 = \hat m^1 n^2 n^3, \quad p_2 = n^1 \hat m^2 n^3, \quad p_3 = n^1 n^2 \hat m^3, \\
&& q_0 = \hat m^1 \hat m^2\hat m ^3, \quad q_1 = - n^1 \hat m^2 \hat m^3, \quad q_2 =  -\hat m^1 n^2 \hat m^3, \quad q_3 =-  \hat m^1 \hat m^2 n^3.\nonumber
\eea
This is in agreement with the general expression (\ref{wrapNum_gen}).

\end{appendix}

\end{document}